\PassOptionsToPackage{pdfpagelabels=true}{hyperref} 
\documentclass[a4paper,fleqn,usenatbib]{mnras}

\usepackage{newtxtext,newtxmath}

\usepackage[T1]{fontenc}
\usepackage{ae,aecompl}

\usepackage{graphicx}	
\usepackage{amsmath}	
\usepackage{color}
\usepackage{fancybox}
\usepackage[utf8]{inputenc}
\DeclareUnicodeCharacter{2609}{\sun}
\usepackage{rotating}
\usepackage{import}
\usepackage{pdfpages}
\usepackage{amsmath}
\usepackage{xcolor}
\usepackage{hyperref}
\usepackage{verbatim} 
\usepackage{CJK}
\usepackage{etoolbox}
\usepackage{mathtools}
\usepackage[justification=centering]{caption}
\usepackage{adjustbox}

\newcommand{\printfnsymbol}[1]{%
  \textsuperscript{\@fnsymbol{#1}}%
}

\usepackage[switch,columnwise]{lineno}

\newcommand{\JL}[1]{\textcolor{black}{#1}}
\newcommand{\rev}[1]{\textcolor{black}{#1}}

\begin{document}

\title[Deep Learning Black Hole Mass]{AGNet: Weighing Black Holes with Deep Learning}


\author[Lin et al.]{
Joshua Yao-Yu Lin,$^{1, 2, 3}$\thanks{equal contribution} \thanks{E-mail: yaoyuyl2@illinois.edu (JYYL), snehjp2@illinois.edu (SP), xinliuxl@illinois.edu (XL)} 
Sneh Pandya,$^{1,2}$\footnotemark[1]
Devanshi Pratap,$^{4,2}$
Xin Liu,$^{5,2}$
\newauthor
~Matias Carrasco Kind,$^{2,5}$
Volodymyr Kindratenko$^{2,4,6}$\\
$^{1}$Department of Physics, University of Illinois at Urbana-Champaign, 1110 West Green Street, Urbana, IL 61801, USA \\
$^{2}$National Center for Supercomputing Applications, 605 East Springfield Avenue, Champaign, IL 61820, USA \\
$^{3}$Center for Computational Astrophysics, Flatiron Institute, Simons Foundation, 162 5th Ave, New York, NY 10010, USA \\ 
$^{4}$Department of Computer Science, University of Illinois at Urbana-Champaign, 201 North Goodwin Avenue, Urbana, IL 61801, USA \\
$^{5}$Department of Astronomy, University of Illinois at Urbana-Champaign, 1002 West Green Street, Urbana, IL 61801, USA \\
$^{6}$Department of Electrical and Computer Engineering, University of Illinois at Urbana-Champaign, 306 North Wright Street, Urbana, IL 61801, USA\\
}

\date{Accepted XXX. Received XXX; in original from XXX}

\pubyear{2021}


\label{firstpage}
\pagerange{\pageref{firstpage}--\pageref{lastpage}}
\maketitle

\begin{abstract}

Supermassive black holes (SMBHs) are \JL{commonly} found at the centers of most massive galaxies. Measuring SMBH mass is \JL{crucial} for understanding the origin and evolution of SMBHs. \JL{Traditional approaches, on the other hand, necessitate the collection of spectroscopic data, which is costly. We present an algorithm that weighs SMBHs using quasar light time series information, including colors, multiband magnitudes, and the variability of the light curves, circumventing the need for expensive spectra.} We train, validate, and test neural networks that directly learn from the Sloan Digital Sky Survey (SDSS) Stripe 82 light curves for a sample of $38,939$ spectroscopically confirmed quasars to map out the nonlinear encoding between SMBH mass and multi-\JL{band} optical light curves. We find a 1$\sigma$ scatter of 0.37 dex between the predicted SMBH mass and the fiducial virial mass estimate based on SDSS single-epoch spectra, which is comparable to the systematic uncertainty in the virial mass estimate. Our results have direct implications for more efficient applications with future observations from the Vera C. Rubin Observatory. Our code, \textsf{AGNet}, is publicly available at
 {\color{red} \url{https://github.com/snehjp2/AGNet}}.

\end{abstract}



\section{Introduction}




Supermassive black holes (SMBHs) with masses $\sim10^{5}M_{\odot}$--$10^{10}M_{\odot}$ are commonly observed at the centers of most massive galaxies \citep[e.g.,][]{KormendyHo2013}. 
%
%
Quasars are accreting SMBHs whose central region outshines the rest of the combined luminosity of the stars in the host galaxy with emission across all wavelengths, providing a window to study how a SMBH grows with time \citep{Soltan1982}. 
Active SMBHs with masses $\sim10^{10}M_{\odot}$ powering the most luminous quasars have formed when the universe was less than a Gyr old after the Big Bang \citep[e.g.,][]{Wu2015,Wang2021}. Understanding how SMBHs formed so quickly is an outstanding problem in astrophysics \citep{Inayoshi2020}. Quasars may also offer a unique ``standard candle'' to study the expansion history of the universe to understand the nature of Dark Energy -- arguably the biggest mystery in contemporary astrophysics \citep[e.g.,][]{King2014,Lusso2017,Dultzin2020}. 

Measuring SMBH mass is important for understanding the origin and growth of quasars \citep{Shen2013}.
However, traditional SMBH weighing methods require spectroscopic data which is highly expensive to gather; the existing $\sim$1,000,000 masses represent $\gtrsim$ 20 years' worth of state-of-the-art community efforts \citep[e.g.,][]{Shen2011,2020ApJS..249...17R}. The Legacy Survey of Space and Time (LSST) at the Vera C. Rubin Observatory \citep{Ivezic2019} will discover $\sim10^8$ new quasars\footnote{\url{https://www.lsst.org/sites/default/files/docs/sciencebook/SB_10.pdf}}, which would not be feasible to weigh with traditional methods. Therefore, a much more efficient approach is needed to maximize LSST SMBH science. \JL{\citet{Burke2021} used optical continuum variability observed in 67 active galactic nuclei and found a correlation between characteristic variability timescale and the SMBH mass. \citet{mchardy2006active} also had similar finding with X-ray variations. While these methods shows promising hints, to fully model each individual SMBH observed in the future could be challenge due to several reasons: (1) The encoding is likely to be highly nonlinear and may be difficult to fully model using traditional human-engineered statistics, (2) the number of observed AGN would be huge.}

Hence, we present a new approach to potentially solve the efficiency problem of SMBH mass estimates based on Deep Learning (DL). DL is an outstanding tool in the current age of astroinformatics and has been applied for many applications in astrophysics, including solar flare forecast \citep{Huang2018}, star-galaxy classification \citep{Kim2017}, large scale structure \citep{Buncher2020}, galaxy surface-brightness-profile fitting \citep{Tuccillo2018}, transient detection \citep{Cabrera-Vives2017}, supernova classification \citep{Charnock2017}, strong gravitational lens detection \citep{Lanusse2018}, gravitational-wave detection \citep{George2018a}, cosmological parameter inference from weak gravitational lensing \citep{Ribli2018}, and image deblending and classification \citep{Burke2019}, just to name a few.  Recently, DL has been employed to classify quasars and predict their cosmological redshifts \citep{Pasquet-Itam2018} using light curve data from the Sloan Digital Sky Survey \citep[SDSS;][]{York2000}. \citet{tachibana2020deep}  has utilized Recurrent Autoencoders (RAE) to extract features from the Catalina Real-time Transient Survey and simulated data to model the quasar variability using single-band light curves, and also searched for potential correlations between the RAE derived features and the physical properties of the sample. \JL{\citet{park2021inferring} used Bayesian neural networks to infer black hole proprieties using simulated AGN light curves.} \citet{LinPandya2020} has applied traditional machine learning techniques in photometric redshift and SMBH mass estimation for SDSS quasars. 

In \citet{LinPandya2020}, the authors used features that are directly using the fitting function from Damped Random Walk parameters, and passed those features to a machine learning algorithm such as a K-Nearest-Neighbors and a multilayer perceptron (KNN and MLP). While the pipeline performs on estimating quasar mass,
the pipeline does not directly use the photometric light curve data for quasar mass estimation. This motivated the further investigation as to whether a hybrid neural network architecture that takes the full photometric light curve
as input data could estimate quasar mass.

In this paper, we introduce \textsf{AGNet}, a hybrid, deep neural network used to estimate SMBH mass and photometric redshift. This work represents the first DL application for predicting quasar's SMBH mass with multi-\JL{band} light curve data. Our overall strategy of using a DL approach is motivated by empirical evidence and theoretical reasons to believe that the quasar light curve encodes physical information about the central SMBH which powers the optical-continuum-emitting accretion disk \citep[e.g.,][]{kelly2009variations,MacLeod2010modeling,Burke2021}. The encoding is likely to be highly nonlinear and may be difficult to fully model using traditional human-engineered statistics. We train deep neural networks in a supervised fashion that directly learn from the light curve data to map out the nonlinear encoding. 
\textsf{AGNet} directly weighs SMBHs using quasar light curves, which are much cheaper to collect for large samples. Our approach will be directly applicable for future, more efficient applications with the LSST, circumventing the need for expensive spectroscopic observations for large samples.  

The paper is organized as follows. We describe our data and methodology of the DL algorithm in \S \ref{sec:method}. We introduce the hybrid deep neural network in \S \ref{sec:nn}. We present our results in \S \ref{sec:result} on summary statistics of quasar SMBH mass estimation. We discuss the implications of our results in \S \ref{sec:discuss}. Finally, we summarize our conclusions and suggest directions for future work in \S \ref{sec:future}.  A concordance $\Lambda$CDM cosmology with $\Omega_m = 0.3$, $\Omega_{\Lambda} = 0.7$, and $H_{0}=70$ km s$^{-1}$ Mpc$^{-1}$ is assumed throughout. We use the AB magnitude system \citep{Oke1974}.

\section{Data and Method}\label{sec:method}

\subsection{Data}

\subsubsection{SDSS Stripe 82 Light Curves}\label{subsubsec:s82}

We adopt multi-\JL{band} photometric light curves from the SDSS Stripe 82, a 2.5 degree wide strip along the Southern Galactic Cap \citep{ivezic07}, as our training and testing data. The SDSS Stripe 82 provides deep photometry in 5 broad optical ($ugriz$) bands. \JL{There are on average 60 epochs of observation on Stripe 82. Observations spanned roughly 2-3 months for a decade of total observation time, providing an effective cadence of time-scales from days to years \citep[e.g.,][]{MacLeod2010modeling, Liao2020}. The photometry used in Stripe 82 is PSF-based as the targets are quasars (i.e. point sources)}.

We gather the Stripe 82 light curves from the publicly available repositories from two main sources. The University of Washington catalog\footnote{The data underlying this article are available in Southern Sample (SDSS Stripe 82)
at \url{http://faculty.washington.edu/ivezic/macleod/qso_dr7/Southern.html}, and can be accessed with \citep{macleod2012description}.
} contains $9,038$ total light curves after cleaning. \JL{Light curves with anomalous data points of mag = $\pm 99$ and unphysical gaps are removed so as to not negatively effect the performance of AGNet.}
The Richards Group LSST Training Set \footnote{The data underlying this article are available in Richards Group LSST Training Set at \url{https://github.com/RichardsGroup/LSSTprep}.} being constructed by researchers at Drexel University contains a total of $84,790$ quasars which is reduced to $38,939$ quasars in Stripe 82. From the Richards Group catalog, we additionally utilize the SDSS point-spread function (PSF) magnitudes that are corrected for galactic extinction for each quasar.




\subsubsection{Virial Black Hole Mass and Spectroscopic Redshift }\label{subsubsec:virialmass}

We take the virial black hole mass estimates from the SDSS DR7 catalog of $\sim$100,000 quasars \citep{Shen2011} and the DR14 Quasar catalog \citep{Paris2018} of $\sim$500,000 quasars \citep{2020ApJS..249...17R} and assume them as the ``ground truth''. These estimates are made from the virial mass method based on SDSS single-epoch spectra \citep{Shen2013}. 


%

\begin{figure}
 \centering
  \includegraphics[width = .5\textwidth]{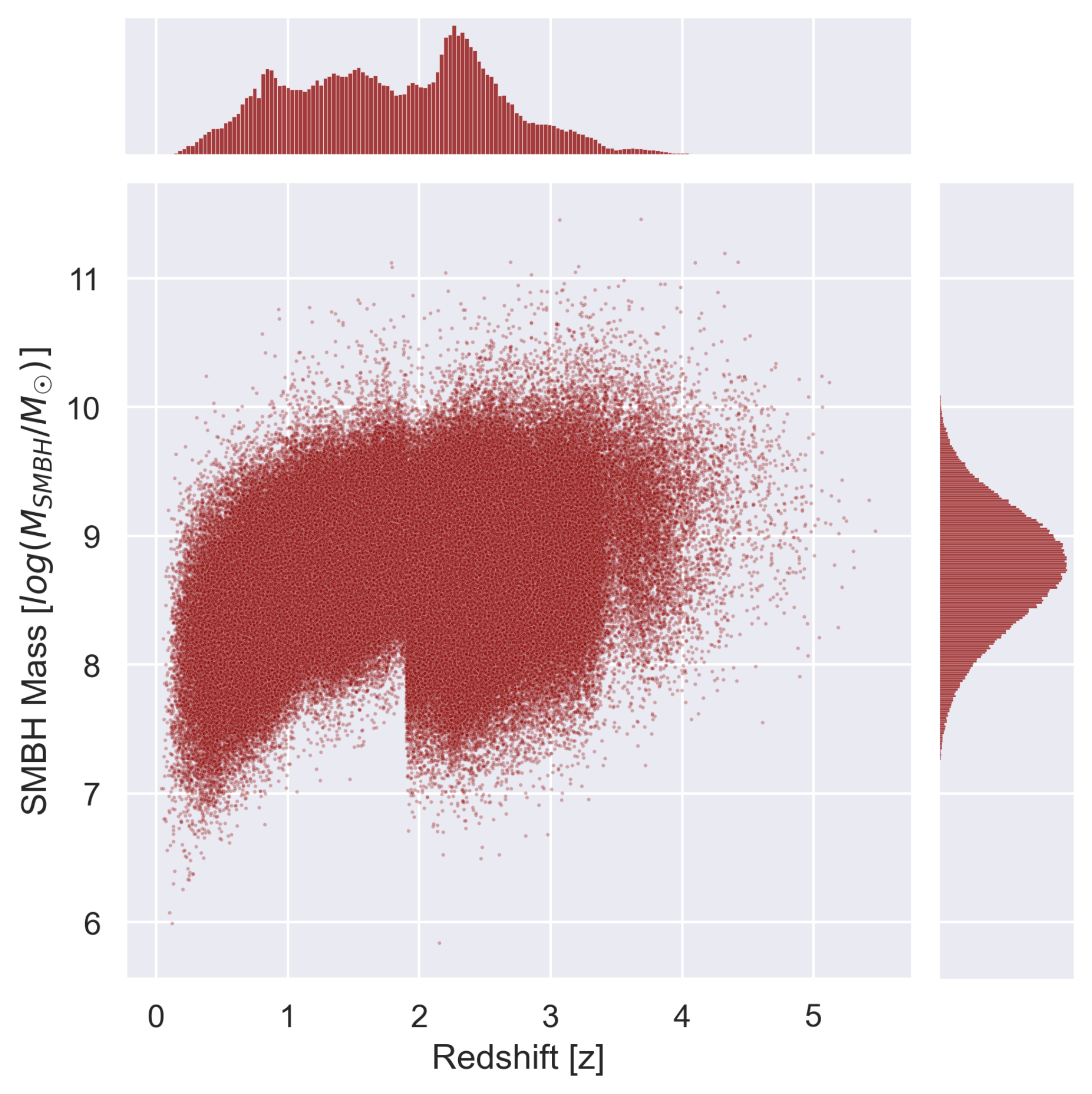}
    \caption{Distribution of $417,618$ cleaned DR14 quasars in mass-redshift space with top and right panels showing quasar redshift and SMBH mass histograms, respectively. Mass is measured in units of $log(M_{SMBH}/M_{\sun})$. Redshifts are provided in the DR7 and DR14 catalogs and cross-matched with redshift values from the Stripe 82 light curves.  Unphysical supermassive black hole masses ($M_{SMBH}<0$) and unreliable estimates with error $>.3$ dex are removed.}
  \label{fig:mass_dist}
\end{figure}

We find that some of the SMBH mass estimates have relatively large uncertainties ($>0.3$ dex) and some unphysical mass measurements, which would negatively affect the performance of AGNet.  In training, we find much improved performance of AGNet when filtering out high uncertainty mass measurements. We clean any unphysical quasar SMBH estimates ($M_{{\rm SMBH}} < 0$) as well as measurements meeting the error criteria.  Eliminating quasars with this criteria reduces our DR14 dataset by $\sim$100,000 objects. A distribution of our cleaned catalog of quasars in mass-redshift space is shown in Figure \ref{fig:mass_dist}.

For estimating quasar's cosmological redshift, we also utilize the spectroscopic redshift from the SDSS DR14 Quasar catalog as the ground truth.  We also find that K-corrected $i$-band magnitude ($M_{i}$), which is available in SDSS DR7 and DR14 \citep{SDSSDR7, Paris2018}, is a useful feature in estimating SMBH mass. \JL{$M_{i}$ is not used as a feature when predicting SMBH redshift as redshift is used in the calculation of $M_{i}$}. The K-correction provides spectroscopic information of the underlying quasar in its rest frame which may also encode (secondary) information of the underlying SMBH mass because of the correlation between quasar redshift and SMBH mass in a flux-limited sample. This relationship correlation between redshift and SMBH mass is clearly seen in Figure \ref{fig:mass_dist}.


\subsection{Data Preprocessing}



\subsubsection{Data Matching}\label{subsubsec:data_matching}

We use \texttt{astropy} to match the two data sets given a 0.5 arcsec tolerance and verify by matching redshift values between the Stripe 82 and DR14 data sets.
In matching our light curve catalog of $\sim$40k light curves, and the cleaned DR14 catalog of $\sim$400,000 quasar virial mass estimates, we find that $\sim$10,000 quasars do not have virial mass estimates in the DR14 catalog.  In addition to the cleaning applied on the DR14 catalog, this brings the final dataset to $\sim$20,000 quasars.

\subsubsection{From Light Curves to Images}


 We aim to use convolutional neural networks (CNNs), which have been shown to have state-of-the-art performance on images, to encode the information from the light curves. While the light curves are not naturally interpreted as 2D images, \citep{Pasquet-Itam2018} has shown promising results on using CNNs on light curves for redshift estimation. To transform the light curve data into images, we take the Stripe 82 photometric light curves as input and the black hole masses as labels. The Stripe 82 light curves have 5 bands (u, g, r, i, z) and span over 3000 MJD, so we transform the light curve data into “images” by putting the 5-band light curves into a $5\times 3340$ numpy array and taking the MJD as an integer using rounding. \JL{To match the dimensionality properly, the arrays are filled with zeros when no data (observations) is present. This is known as zero padding.} The data would need to have zero padding when the light curve data are in the gaps. We then interpolate it into $224\times 224$ numpy array images. For using the pretrained model, we stack over the same $224\times 224$ images three times to fill the RGB channel.

\subsubsection{Feature Extraction from Light Curves}

In addition to using the deredded ugriz bands as features, we compute the colors $(u-g, g-r, r-i, i-z, z-u)$ as features for a quasar.  For the RichardsGroup light curves, we adopt the PSF magnitudes and compute the colors from these.  For the University of Washington light curves, we take the mean of each band observation across the observational epoch and compute the colors from the mean. The colors are more robust features than the individual bands in that they provide more information regarding the quasar’s spectral energy distribution (SED) and temperature. To standardize the effect of our features in training we apply the \texttt{scikit-learn StandardScaler} which removes the mean and scales the data to unit variance. \JL{For SMBH mass estimation, we add additional features including best-fit damping time scale $\tau$, best-fit driving amplitude of short-term fluctuations $\sigma$ as well as K-corrected absolute i band magnitude $M_i$. }

\subsubsection{Train, Validation, and Test Set Splitting}

\JL{We split our baseline data set into an $70\%$ training, $15\%$ validation, and $15\%$ testing set using the \texttt{sklearn train test split}. It is best-practices in machine learning to train a model on the bulk of the available data using the training set. During training, the model can be continuously evaluated on the validation set to benchmark performance and tune the hyperparameters of the model. The validation performance is also a good indicator if the model is \emph{overfitting}, the phenomena where a model will learn features of the training set that are exclusive to the training distribution and will not aid the models performance in real-world applications. The testing set is used for a final, unbiased evaluation of the model's performance.}

\subsubsection{Underlying Spectral Energy Distribution}

A quasars SED provides information of the quasar's energy output as a function of its emitted wavelength. 
A quasar's observed SED is primarily determined by the intrinsic luminosity, accretion rate, and redshift \citep{Elvis1994,Richards2006}. It also contains secondary information about the black hole mass through the dependence on luminosity and accretion rate.
By providing features (bands, colors) that are indicative of the underlying SED, we feed the neural network with information to infer the SMBH mass. 

In analysis of the principal components of our data in \S \ref{sec:pca} , we see artifacts of the SED in our features.  The $u$-band has the shortest wavelength of the five bands, and therefore also the largest variability amplitude \citep[e.g.,][]{diClemente1996,Ulrich1997}.  The $z-u$ color provides the largest difference in wavelength and has the largest color gradient. These characteristics are most sensitive to the shape of the quasar's SED \citep{dunlop2003quasars}.


\begin{figure*}
 \centering
  \includegraphics[width=\textwidth]{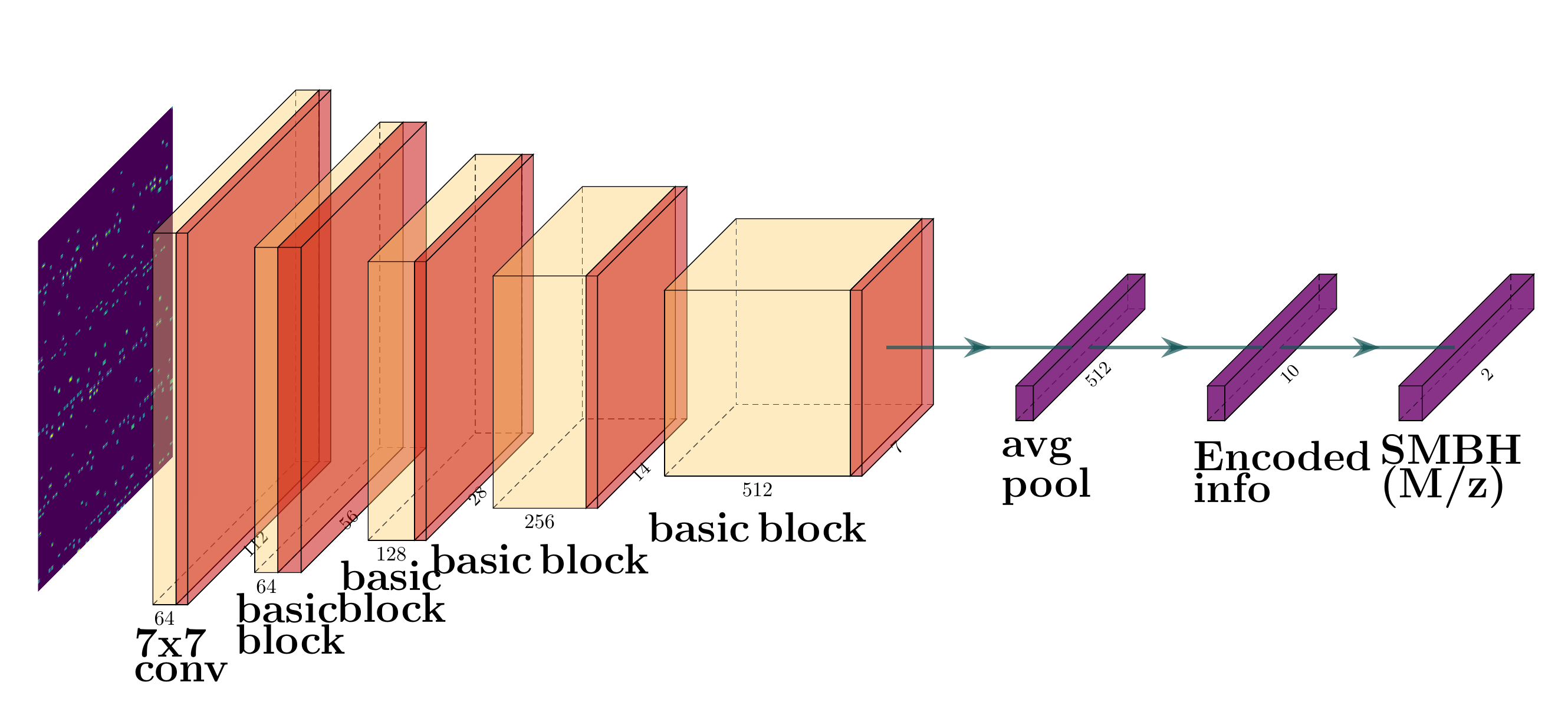}
    \caption{Convolutional Neural Network Architecture. A basic ResNet block is composed of two layers of 3x3 convolutions, subsequently applying a \texttt{batchnorm} transformation and \texttt{ReLU} as activation. We modified the last layer of the ResNet18 so it outputs the parameter of our interest (quasar SMBH mass/redshift).}
  \label{fig:CNN}
\end{figure*} 

\begin{figure*}
 \centering
  \includegraphics[width=\textwidth]{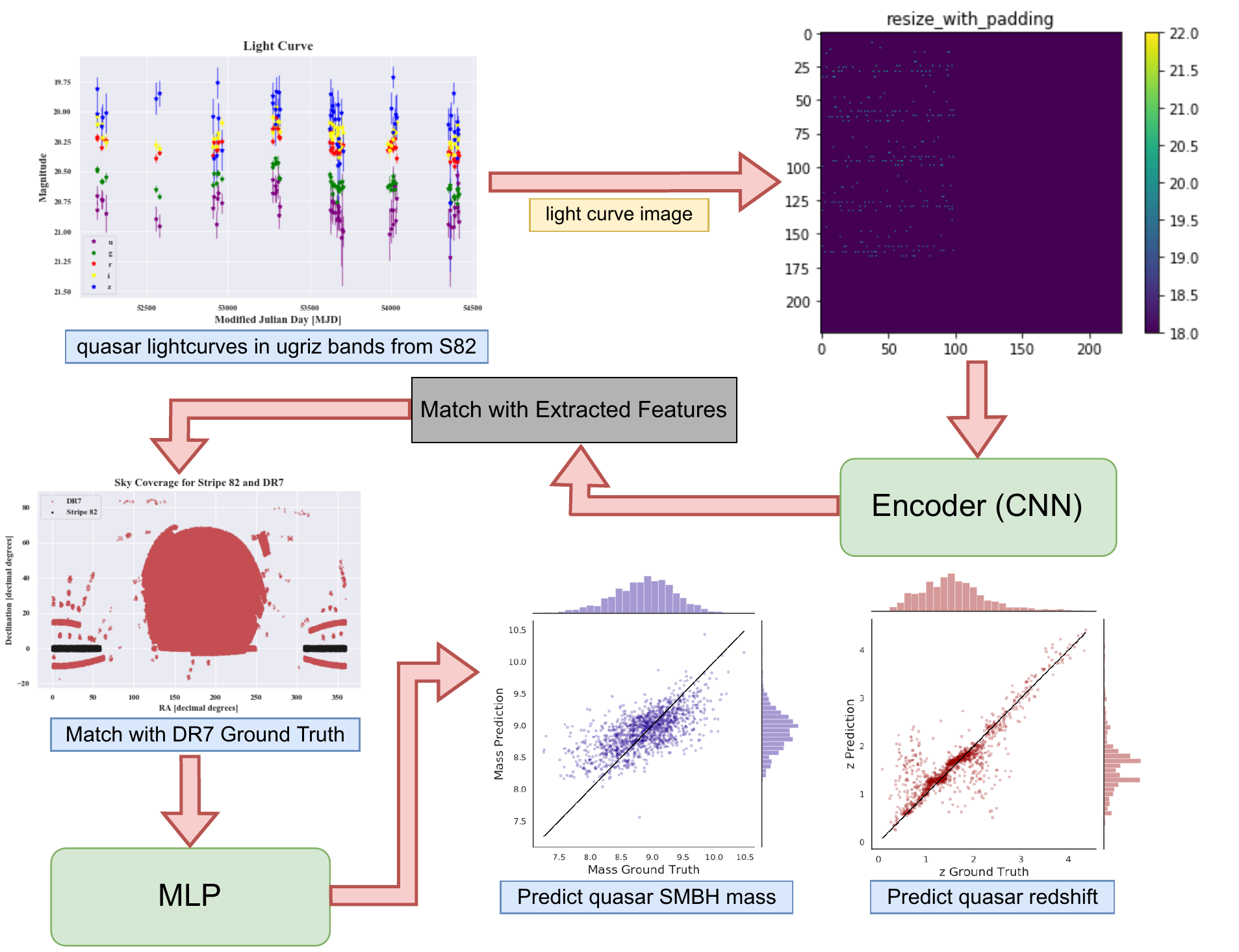}
    \caption{AGNet: A hybrid combination of a MLP and CNN.}
  \label{fig:AGNet pipeliine}
\end{figure*} 

\section{Deep Neural Network}\label{sec:nn}

\subsection{Introduction to Neural Networks}

\JL{The foundation of modern deep learning and machine learning is the \emph{deep neural network} (DNN), a computational model that is inspired by biological neurons. Individual neurons in general are parameterized by a matrix of weights and a bias vector. These networks and their intrinsic parameters can be refined to learn structure of data or predict some output through the process of training via gradient descent. A function known as the \emph{loss function} is responsible for how the network learns, and is chosen according to the machine learning task at hand (regression vs. classification) and the quality of the data. Finding the global minima of the loss function's associated \emph{loss landscape} corresponds to the best-performance achieved by the DNN. Certain architectures of DNNs have been shown to excel at specific tasks \citep{alexnet}. For example, CNNs are excellent candidates in analyzing image data, whereas the more simple multilayer perceptron (MLP) is sufficient when dealing with vectorized data. The first stage of the AGNet pipeline utilizes a ResNet18 architecture, which is a pretrained CNN, to output an encoded representation of the light curves. The encoded light curve features from the ResNet18 are combined with other hand-selected features and fed into a MLP in the second stage of the pipeline. The MLP is then trained on and outputs the SMBH mass.}

\JL{\emph{Transfer learning} is a popular technique used in applying deep learning algorithms. Many state-of-the-art deep learning architectures are trained and benchmarked on large datasets like ImageNet, containing 14 million images. These pretrained models' weights have already been optimized to recognize features in images and therefore should perform well on complex tasks without the computational efficiency needed to train them from scratch. The models with their pretrained weights can be readily accessed through popular deep learning libraries. In this work and others which utilize transfer learning, the output layer weights are retrained for the task at hand.}

\subsection{Convolutional Neural Networks}
\label{sub: CNN}

\JL{CNNs can be decomposed into three main components: a convolutional layer, pooling layer, and fully-connected layer. The convolutional layer functions by passing a filter or kernel over the image, which produces a representation of the image that is sensitive to features defined by the kernel. In this way, convolutional layers are able to efficiently extract features from an image. The pooling is used to 'fill in the gaps' after the convolution operation. The pooling layer typically assigns the value to missing pixels to be the mean of the neighboring pixels. The fully connected layer at the end allows the flexible application of CNNs to regression and classification tasks. 
}

\JL{Deep residual neural networks, or ResNets, are a successful architecture \citep{he2016deep} that introduces the \emph{residual block}. The residual block features all the aforementioned components of typical CNNs with the introduction of \emph{skip connections}. Skip connections introduce alternative passes in the information-flow through the network and have been shown effective in avoiding the \emph{vanishing gradient problem} \citep{he2016deep}. If a network encounters vanishing gradients, it gets stuck in a local minima or saddle point of the loss landscape and ceases to learn more information. It has been shown that neural networks with skip connections can better approach the minimum of highly non-convex loss functions with smoother loss surface \citep{li2018visualizing}. This further motivates the use of pretrained ResNets in scientific research.}



In recent years, CNNs have shown promising results on classifying or extracting parameters from astrophysical images \citep{hezaveh2017fast,khan2019deep,lin2020feature,wu2020predicting}. While most CNNs are designed primarily for 2-dimensional images, we found that they could be applied appropriately to our light curve data after reshaping the 1-dimensional data into 2-dimensional images with zero padding \JL{as shown in Figure \ref{fig:CNN}}. \JL{The pretrained ResNet18 is modified and retrained in its last two layers to output 10 encoded light-curve features. The addition of the fully-connected layers after the CNN allows the desired parameter output.} \JL{The first branch of the} neural network is a feature extractor, wherein earlier layers detect low-level features (e.g. each magnitude measured at individual MJD) and later layers detect high-level features (e.g. magnitude correlation across larger MJD). 


\subsection{Fully-connected Neural Networks}

We use fully-connected neural networks, also known as multilayer perceptrons (MLPs), as our basic framework to predict SMBH mass and redshift. Our MLP architecture for SMBH mass prediction is a deep neural network with a 12-neuron input layer followed by 5 hidden layers. Our architecture for redshift is identical with exception of the input size (10-neuron input layer). The mass estimation pipeline uses information from the SDSS bands, colors, redshift, and $M_i$ as features, and just the SDSS magnitudes and colors as features in predicting quasar redshift. Details of the input features are listed at Table \ref{tab:summary_stats}.

We empirically determined the best set of hyperparameters for our MLP. We use ReLU activation on all neurons \citep{agarap2018relu} after each fully connected layer.

\subsection{AGNet: A Hybrid Combination of CNN and MLP}

A good representation of our data will contain information to estimate the SMBH's parameters. To maximize the information we could extract from the light curves, we merge the information from our feature extractors and the encoded representation from CNN as shown in Figure \ref{fig:AGNet pipeliine}. AGNet is a hybrid combination of CNN and MLP that utilizes advantages of both architectures. In AGNet, the CNN acts as an encoder to generate a representation that could potentially provide more robust information from the light time series, while the manually extracted features provide adequate spectral information to estimate the SMBH mass. 

\JL{To train the CNN as an encoder, we first use the CNN to train an end-to-end prediction for redshift and SMBH masses directly from the the light curve images, as mentioned in section \ref{sub: CNN}. Once the training is done, we extract the features at the last hidden layer and concatenate with other physical features.}

\subsection{Loss Function and Hyperparameters}


We train our neural network with gradient descent based \texttt{AdamW} optimizer, an adaptive learning rate optimization algorithm \citep{kingma2014adam, loshchilov2017decoupled}, with default learning rate set as $0.001$ for 50 epochs of training. \JL{The learning rate parameter dictates the step-size of the neural network in the loss-landscape throughout training.} Small learning rates ($.005 < lr < .001$) were found to work best. Instead of using a standard mean squared error (MSE) loss, also known as $\chi^2$ and commonly used for fitting astrophysical data, we optimize the MLP with SmoothL1 loss to minimize the effect of outlier SMBH masses in training, and encourage appropriate mass estimations at both the high ($> 10^9 M_{\odot}$) and low-end ($< 10^8 M_{\odot}$) of the ground truth SMBH mass distribution. The SmoothL1 loss function is given as:

\begin{equation}\label{eq:loss}
\mathcal{L} = 
\begin{cases}
                $0.5$(x-y){^2} , & \lvert x-y \rvert $ < $ $1$  \\
                \lvert x-y \rvert - $0.5$ , & \text{otherwise}
                
\end{cases}
\end{equation}
where $x$ is the network prediction and $y$ is the ground truth value.

The MLP operates with $\sim$30,000 trainable parameters.



In this work, we use \texttt{PyTorch} \citep{NEURIPS2019_9015}, a Python3 deep learning library to build our neural network.  Model training was done on an individual workstation utilizing an Nvidia 1080Ti GPU and the HAL cluster at the National Center for Supercomputing Applications (NCSA) \citep{10.1145/3311790.3396649}.







\section{Results}\label{sec:result}

To evaluate our models, we use the $R^2$ score metric, also known as the coefficient of determination, which is a function of root mean square error (RMSE) as a metric to evaluate the network performance. RMSE is defined as the average square of the difference vector between ground truth and prediction.  Table \ref{tab:summary_stats} lists the results of our network performance. 
The $R^2$ score is defined as:
\begin{equation}
    1-\frac{{\rm variance~of~data}}{{\rm mean ~of~squared~ residuals}},
\end{equation}
which is a measure of the difference between the prediction and the ground truth. 


\subsection{Network Performance on Redshift}\label{sec:zresults}


Following \citet{Pasquet-Itam2018}, we compare our network performance on redshift estimation. We use ugriz bands and colors as features for predicting redshift following the architecture outlined above.  Our best AGNet performance gives a RMSE = $0.373$ and a $R^2 = 0.724$, which is the same as the MLP and better than CNN, showing that AGNet is performing well on the redshift task compared with other models, and our result on redshift is also comparable to \citet{Pasquet-Itam2018}. Results can be found in Table \ref{tab:summary_stats}.

\begin{figure}
 \centering
  \includegraphics[width=85mm]{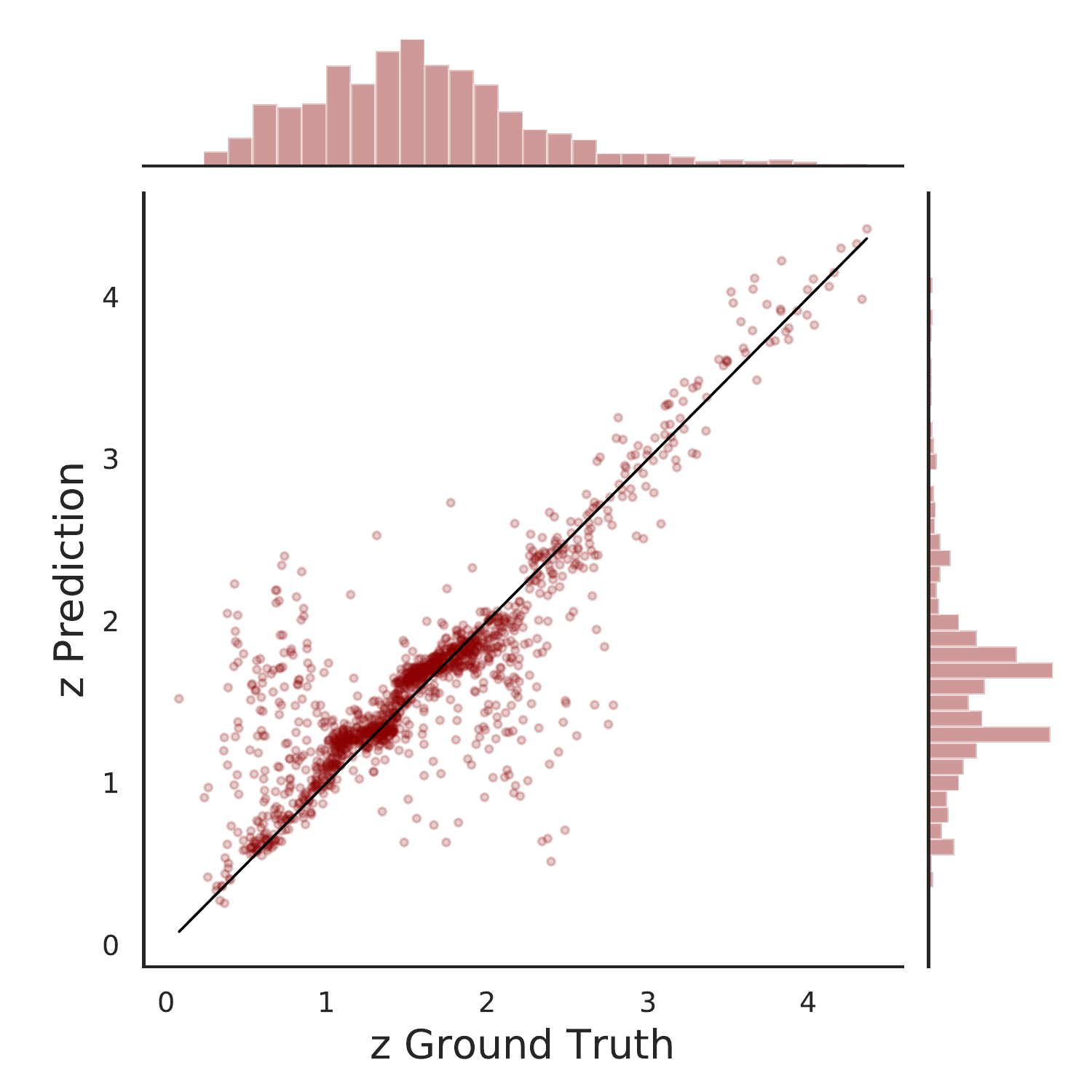}
    \caption{AGNet predictions for quasar SMBH redshift. Ground truth redshifts are shown by 1:1 black line with network predictions as the scatter. Trained on over $27,000$ quasars and tested on roughly $5,800$ quasars.}
  \label{fig:redshift results}
\end{figure}

There are noticeable spikes in redshift estimation at $z\sim0.9$ and $z\sim2$.  These are due to degeneracies in the colors of the quasars \citep[e.g.,][]{YangQ2017} as shown in Figure \ref{fig:redshift results}. \rev{We also noticed that the same scores are obtained with and without the light curve image, so the redshift determination is completely dominated by the colours and the single band magnitudes. We trained AGNet with only colors and multi-bands, and we get (RMSE: 0.396, r2 score 0.689) for bands only, and (RMSE: 0.441, r2 score 0.617) for colors only -showing that AGNet performs better when both colors and multi bands are used as features for redshift estimation. We found no further improvement when the light curve information is provided - showing that 1) the existing pipeline for redshift estimation relies on color and multibands formation rather than variability or 2) the encoded representation of the light curve does not capture any existing variability that may correlate with redshift.}

\subsection{Network Performance on Black Hole Mass}\label{sec:massresults}

  Our best performance of AGNet gives a RMSE = $0.371$ and a $R^2 = 0.728$, which outperforms both the MLP and CNN, as well as KNN \S \ref{sec:KNN} . Our results also achieve the mean value of 1-sigma errors in the catalog.  


All of the results presented here are from the predictions of the test set to prevent over-fitting from the training set.

\begin{figure}
 \centering
  \includegraphics[width=85mm]{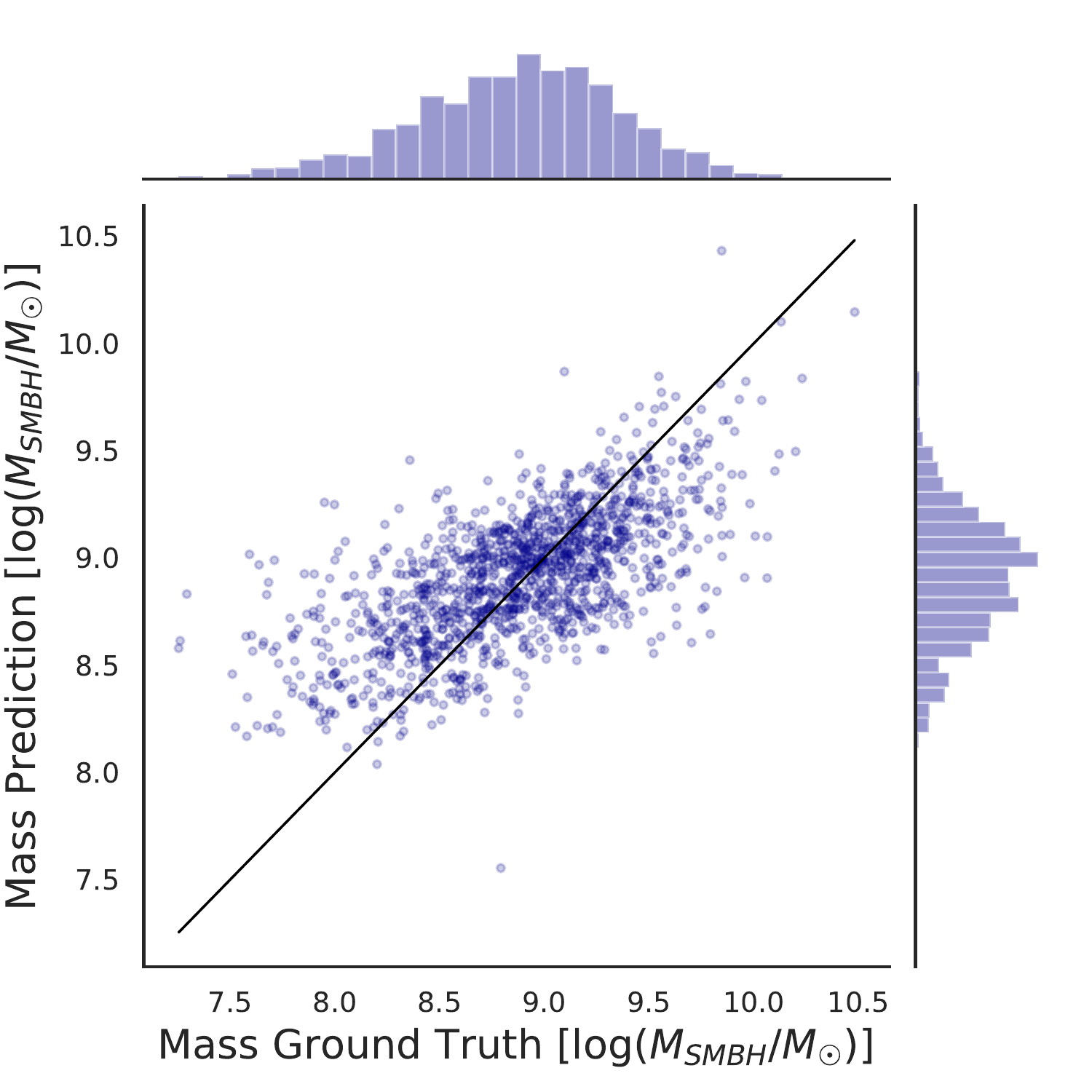}
    \caption{AGNet predictions for quasar SMBH mass. Ground truth masses are shown by 1:1 black line with network predictions as the scatter. Trained on over $27,000$ quasars and tested on roughly $5,800$ quasars.}
  \label{fig:mass results}
\end{figure} 

Our best CNN performance gives a RMSE = $0.481$ and a $R^2 = 0.040$. Our best MLP performance gives a RMSE = $0.385$ and a $R^2 = 0.384$. More statistics can be found in Table \ref{tab:summary_stats}.

\JL{We also found that the model consistently predicts lower masses for quasars with masses larger than $10^{9.3} M_{\odot}$, and consistently larger masses for quasars with masses below $10^{8.5} M_{\odot}$.} \rev{
We use spectroscopic redshift (spec-z) for one of the features as our AGNet input when performing SMBH mass estimation. However, in some cases in the future (e.g. era of LSST), we may not have spec-z for our target AGN so we also train an AGNet without spec-z provided. We found that AGNet performs the best when spec-z is provided. We have color-coded the SMBH mass prediction with $M_i$ (given by the catalog).There exists a bias (that exists across all training/validation/testing sets). However, we still find in our analysis that AGNet also takes into account other information from colors and other bands, as well as variability of the light curve.}

\begin{table*}
\centering
\begin{tabular}{|p{3cm}||p{5cm}|p{3cm}||p{1.25cm}|p{1.25cm}|}
 \hline
 \multicolumn{5}{|c|}{Summary Statistics} \\
 \hline
 ML algorithm & Inputs & Parameters & RMSE & $R^2$\\
 \hline
 AGNet  & light curve image + colors and bands & redshift (z) & 0.373 & 0.724 \\
 CNN & light curve image & redshift (z) & 0.521 & 0.451 \\
 MLP & colors and bands & redshift (z) & 0.373 & 0.724\\
 KNN   & colors and bands &  redshift (z) & 0.442 & 0.603\\
 \hline
 AGNet (w spec-z)& light curve image + colors, $\tau$, $\sigma$, $M_i$, z & SMBH mass & 0.371 & 0.428 \\
 CNN & light curve image & SMBH mass & 0.481 & 0.04 \\
 MLP (w spec-z) & colors, $\tau$, $\sigma$, $M_i$, z & SMBH mass & 0.385 & 0.384\\
 KNN   (w spec-z) & colors, $\tau$, $\sigma$, $M_i$, z & SMBH mass & 0.388  & 0.374\\
 \hline
 AGNet (w/o spec-z)  & light curve image + colors,  $\tau$, $\sigma$, $M_i$, z* & SMBH mass & 0.384 & 0.385 \\
 CNN  & light curve image & SMBH mass & 0.481 & 0.04 \\
 MLP (w/o spec-z) & colors,  $\tau$, $\sigma$, $M_i$, z* & SMBH mass & 0.393 &  0.358\\
 KNN   (w/o spec-z) & colors,  $\tau$, $\sigma$, $M_i$, z* & SMBH mass & 0.396 &  0.350\\
 \hline
\end{tabular}
\caption{Summary Statistics. Above z* denotes predicted redshift values. 'spec-z' denotes spectroscopic redshift. \JL{When spec-z is provided, the best version of AGNet achieves a RMSE of 0.371. When only using photometric information (with out spec-z), the best version of AGNet achieves a RMSE of 0.384.} \label{tab:summary_stats}}
\end{table*}

\section{Discussion}\label{sec:discuss}


In this paper, we have shown that using photometric light curves our AGNet pipeline, which consists of a feature extraction step and neural networks, can provide a fast and automatic way to predict SMBH mass and redshift. We draw the following conclusions:
\begin{itemize}
    \item AGNet is able to learn from information provided from the quasar light time series without expensive spectroscopic spectra.
    \item Our prediction reaches the uncertainty limit of the ground truth data (i.e., systematics in the single-epoch virial estimates), however with better estimates from multi-object reverberation mapping \citep[e.g.,][]{Shen2016, dalla2020sloan,Li_2021}, we could in principle improve the quality of ground truth and hence improve the performance of AGNet.
\end{itemize}
 In the future, our pipeline could serve as a tool to give efficient predictions using photometric light curves only and grant us a glimpse of the underlying SMBH mass distribution of quasars.


\subsection{Principal Component Analysis on Selected Features}\label{sec:pca}

Principal Component Analysis (PCA) is an unsupervised statistical technique used to reduce the dimensionality of data such that the information explaining a certain percentage of the variance is preserved. In our analysis we choose to preserve $95\%$ of the variance.

It is also possible to train on the principal components. This can be beneficial to generalize a model and reduce risk of over fitting. For our purposes, we use PCA as an analysis tool for the features used in the MLP and still train on the original light curves and features.

\begin{figure}
 \centering
  \includegraphics[width=90mm]{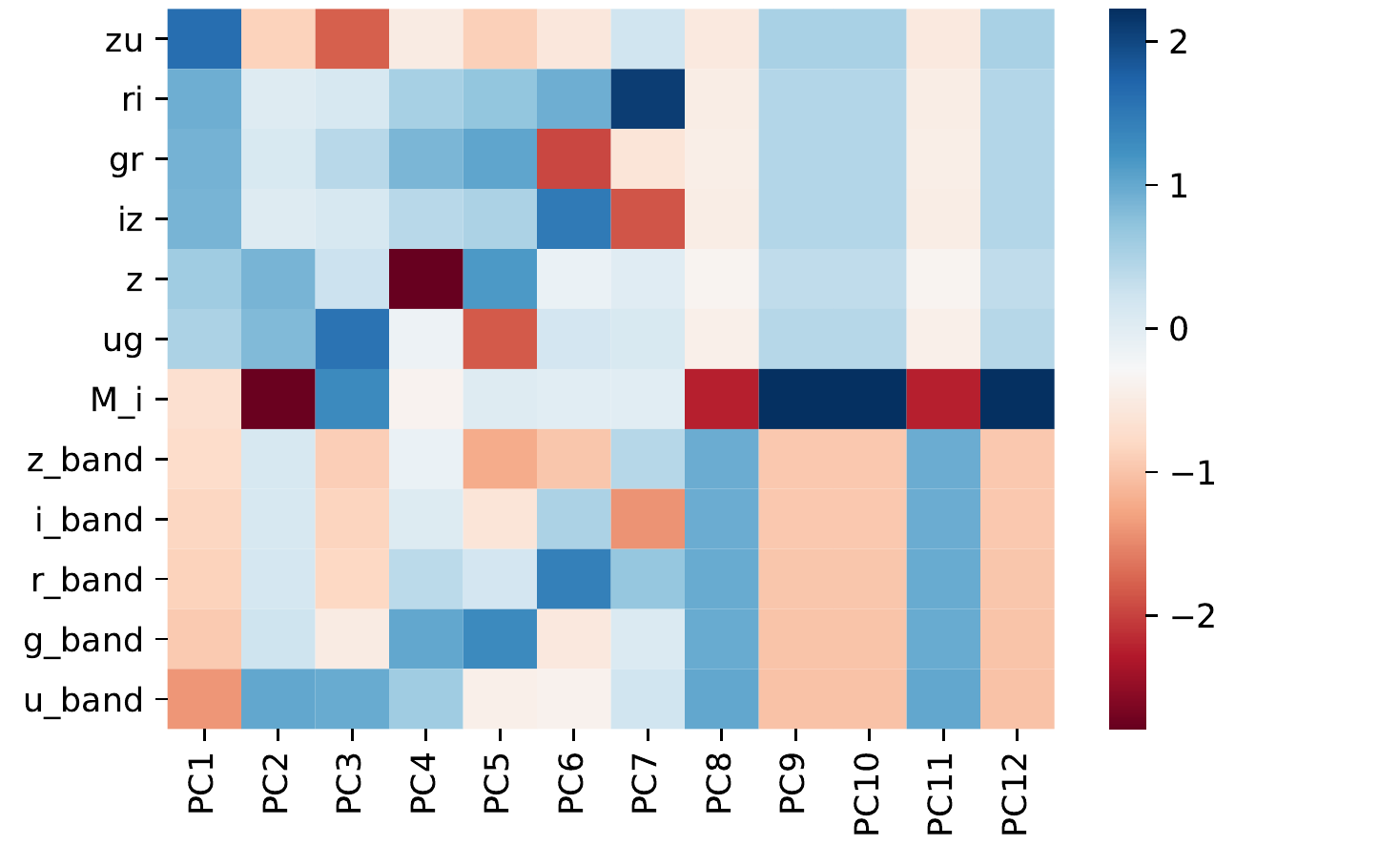}
    \caption{A heatmap showing the correlations between features and principal components.  Positive correlations are shown in shades of blue and negative correlations in shades of red. Correlations are broken down by effect on individual principal componenets.}
  \label{fig:heatmap}
\end{figure}
\begin{figure}
 \centering
  \includegraphics[width=90mm]{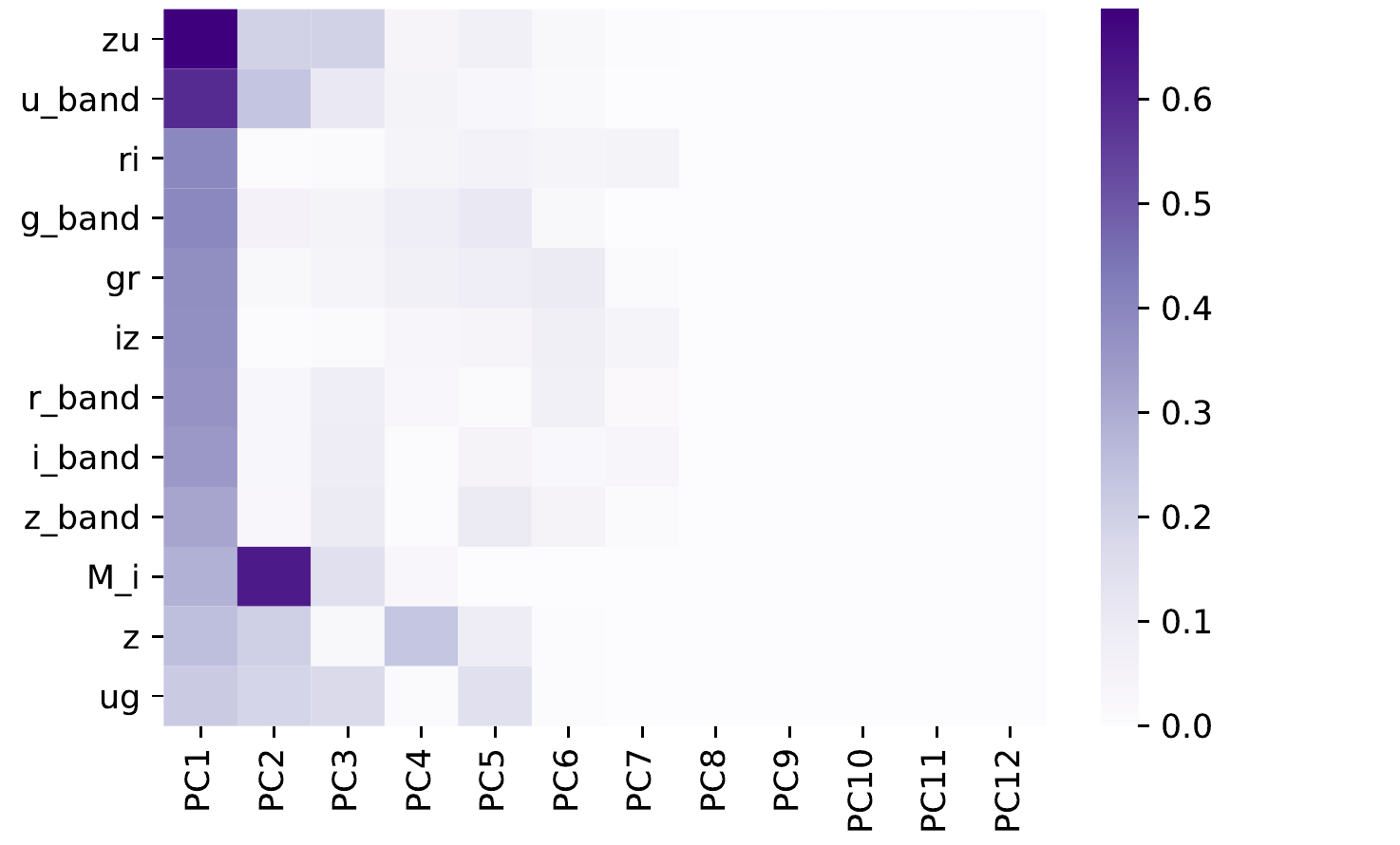}
    \caption{A re-normalized heatmap with the explained variance of each principal component. Darker shades denote that the selected feature explains a large percentage of the variance in a selected principal component.}
  \label{fig:variancemap}
\end{figure}

It is shown that to preserve $95\%$ of the variance that we could cut the dimensionality of our data in half.  Figure \ref{fig:heatmap} and figure \ref{fig:variancemap} show the results of our PCA analysis in features predicting SMBH Mass.  It is shown in Figure \ref{fig:variancemap} that to preserve $95\%$ of the variance, the initial 12-dimensional feature set can be cut in half.  It is additionally shown that the z-u color explains most of the variance in the first principal component, and similarly $M_{i}$ in the second principal component. \rev{Our PCA results indicate that much of the variance of our features comes from $z-u$ color and $M_i$. However, with internal tests with linear regression using z-u and $M_i$ showing that using only those features would not achieve such a performance on SMBH mass estimation, suggesting that AGNet is still able to capture some of the information using other features, including the nonlinearity of the variability from the light curve images.}

 
\subsection{Comparison to Machine Learning KNN Algorithm}\label{sec:KNN}

We now compare AGNet performance to a K-Nearest Neighbors (KNN) algorithm following \cite{Pasquet-Itam2018}. \JL{A KNN algorithm is traditionally used for classification but can also be adapted for regression tasks. The algorithm makes predictions by analyzing the proximity data points have with their \emph{neighbors}. For an arbitrary data point in a regression task, the predicted value (SMBH mass) takes on the average of the values of the neighbors. The number of neighbors considered, \emph{K}, is determined by the user.} We follow the same features and preprocessing as our AGNet implementation, with an ideal K value of $K=19$. \JL{The optimal value of K was found through a simple search.} For quasar redshift, we achieve a RMSE = $0.442$ and for mass we achieve a RMSE = $0.388$ which can be seen in Table \ref{tab:summary_stats}. Using AGNet predicted redshift (z*) values, KNN achieves RMSE = $0.396$, which is comparable to AGNet performance. The KNN performance suggests that spectral and time series features of limited data cannot be extracted as efficiently by traditional machine learning methods. 




\subsection{Future Work}\label{sec:future}



For future research, there are many improvements that can be made. Constructing a streamlined quasar catalog that contains more training data than what was used in this work would be beneficial. We will expand our work with larger data sets in future, such as 
the LSST. Improving the quality of our ground truth SMBH mass measurements by using reverberation mapping measured values would improve the quality of AGNet. Our predictions reach the uncertainty limit of the ground truth data, however with better estimates from reverberation mapping measured masses
we can in principle improve the performance. In terms of model parameters, time series features could be further studied to predict the mass of quasars. We will explore additional time series features for AGNet to learn from using the python library Feature Analysis Time Series (FATS) \citep{nun2015fats}. Uncertainty measurements in redshift and SMBH mass predictions will be included in the future work. We plan on implementing negative log-likelihood loss to quantify uncertainties in our network predictions \citep{levasseur2017uncertainties}. 

\JL{Understanding potential biases in our training set and  how they would effect the predictions could be crucial. \citet{sanchez2021searching} provides a useful analysis and could be potentially helpful for related work in the future.}

This paper establishes feasibility of the DL approach for quasar black hole mass for the first time on multi-band light curves. Given the large volume of data expected from upcoming surveys such as the Vera C. Rubin Observatory, AGNet could provide a pivotal tool in SMBH research.




\section*{Acknowledgement}


We thank Dawei Mu at the National Center for Supercomputing Applications (NCSA) for his assistance with the GPU cluster used in this work and Weixiang Yu for help with the Richards Group LSST Training Set. S.P. and D.P. acknowledge support from the NCSA Students Pushing Innovation Program and the Program Director Olena Kindratenko for guidance. J.Y.Y.L. and X.L. acknowledge support by the NCSA Faculty Fellowship and NSF grants AST-2108162 and AST-2206499. This work utilizes HAL cluster \citep{10.1145/3311790.3396649} supported by the National Science Foundation’s Major Research Instrumentation program, grant No. 1725729, as well as the University of Illinois at Urbana-Champaign. This research was supported in part by the National Science Foundation under PHY-1748958.

Funding for the Sloan Digital Sky Survey IV as been provided by the Alfred P. Sloan Foundation, the U.S. Department of Energy Office of Science, and the Participating Institutions. SDSS-IV acknowledges support and resources from the Center for High-Performance Computing at the University of Utah. The SDSS web site is www.sdss.org.

SDSS-IV is managed by the Astrophysical Research Consortium for the Participating Institutions of the SDSS Collaboration including the Brazilian Participation Group, the Carnegie Institution for Science, Carnegie Mellon University, the Chilean Participation Group, the French Participation Group, Harvard-Smithsonian Center for Astrophysics, Instituto de Astrof\'isica de Canarias, The Johns Hopkins University, Kavli Institute for the Physics and Mathematics of the Universe (IPMU) / University of Tokyo, Lawrence Berkeley National Laboratory, Leibniz Institut f\"ur Astrophysik Potsdam (AIP),  Max-Planck-Institut f\"ur Astronomie (MPIA Heidelberg), Max-Planck-Institut f\"ur Astrophysik (MPA Garching), Max-Planck-Institut f\"ur Extraterrestrische Physik (MPE), National Astronomical Observatories of China, New Mexico State University, New York University, University of Notre Dame, Observat\'ario Nacional / MCTI, The Ohio State University, Pennsylvania State University, Shanghai Astronomical Observatory, United Kingdom Participation Group,Universidad Nacional Aut\'onoma de M\'exico, University of Arizona, University of Colorado Boulder, University of Oxford, University of Portsmouth, University of Utah, University of Virginia, University of Washington, University of Wisconsin, Vanderbilt University, and Yale University.

\section*{Data Availability Statement}

The data underlying this article are available in Southern Sample (SDSS Stripe 82) at \url{http://faculty.washington.edu/ivezic/macleod/qso_dr7/Southern.html}, and can be accessed with \citep{macleod2012description}. Another data underlying this article are available in Richards Group LSST Training Set at \url{https://github.com/RichardsGroup/LSSTprep}.




\bibliographystyle{mnras}

\bibliography{refs} 


\bsp	
\label{lastpage}
\end{document}